\documentclass[journal]{IEEEtran}
\usepackage[latin9]{inputenc}
\usepackage{color}
\usepackage{float}
\usepackage{amsmath}
\usepackage{amssymb}
\usepackage{graphicx}

\makeatletter

\providecommand{\tabularnewline}{\\}
\floatstyle{ruled}
\newfloat{algorithm}{tbp}{loa}
\providecommand{\algorithmname}{Algorithm}
\floatname{algorithm}{\protect\algorithmname}

\ifCLASSINFOpdf
\else
\fi
\usepackage{algorithm}
\usepackage{algpseudocode}

\makeatother

\begin{document}

\title{Deep Cross-Modal Correlation Learning for Audio and Lyrics in Music
Retrieval}

\author{\IEEEauthorblockN{Yi Yu$^{1}$, Suhua Tang$^{2}$, Francisco Raposo$^{3}$\thanks{Francisco was involved in this work during his internship in National Institute
of Informatics (NII), Tokyo.}, Lei Chen$^{4}$} \\
 \IEEEauthorblockA{$^{1}$Digital Content and Media Sciences Research Division, National
Institute of Informatics, Tokyo \\
 $^{2}$Dept. of Communication Engineering and Informatics, The University
of Electro-Communications, Tokyo \\
$^{3}$Instituto Superior Técnico, Universidade de Lisboa, Lisbon\\
$^{4}$Department of Computer Science and Engineering, Hong Kong University
of Science and Technology}}
\maketitle
\begin{abstract}
Deep cross-modal learning has successfully demonstrated excellent
performances in cross-modal multimedia retrieval, with the aim of
learning joint representations between different data modalities.
Unfortunately, little research focuses on cross-modal correlation
learning where temporal structures of different data modalities such
as audio and lyrics are taken into account. Stemming from the characteristic
of temporal structures of music in nature, we are motivated to learn
the deep sequential correlation between audio and lyrics. In this
work, we propose a deep cross-modal correlation learning architecture
involving two-branch deep neural networks for audio modality and text
modality (lyrics). Different modality data are converted to the same
canonical space where inter modal canonical correlation analysis is
utilized as an objective function to calculate the similarity of temporal
structures. This is the first study on understanding the correlation
between language and music audio through deep architectures for learning
the paired temporal correlation of audio and lyrics. Pre-trained Doc2vec
model followed by fully-connected layers (fully-connected deep neural
network) is used to represent lyrics. Two significant contributions
are made in the audio branch, as follows: i) pre-trained CNN followed
by fully-connected layers is investigated for representing music audio.
ii) We further suggest an end-to-end architecture that simultaneously
trains convolutional layers and fully-connected layers to better learn
temporal structures of music audio. Particularly, our end-to-end deep
architecture contains two properties: simultaneously implementing
feature learning and cross-modal correlation learning, and learning
joint representation by considering temporal structures. Experimental
results, using audio to retrieve lyrics or using lyrics to retrieve
audio, verify the effectiveness of the proposed deep correlation learning
architectures in cross-modal music retrieval.\end{abstract}


\section{Introduction}

\begin{figure}
\centering

\includegraphics[width=4cm]{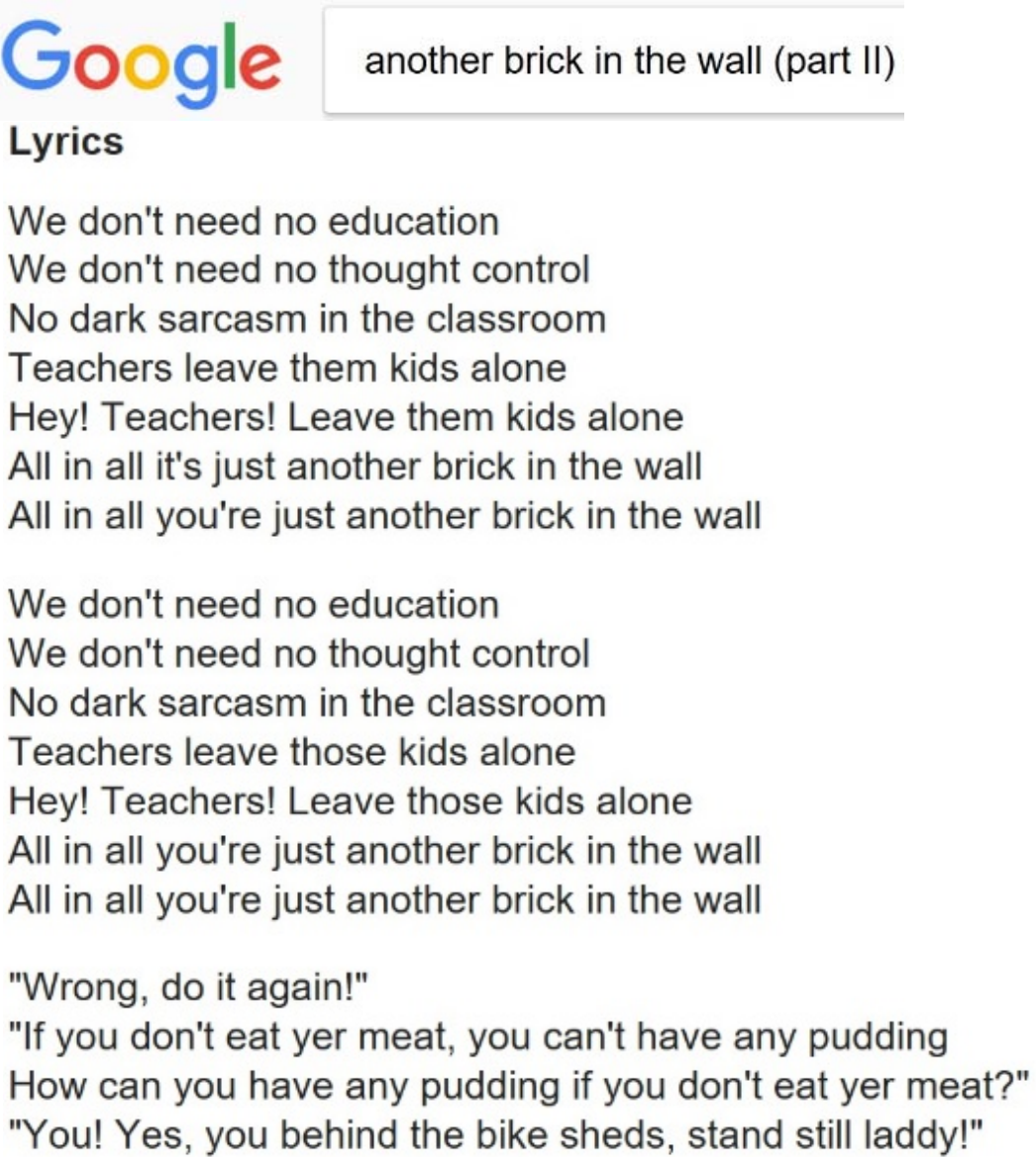}\caption{\label{fig:lyrics.pdf} Google lyrics for song title ``another brick
in the wall (part II)''.}
\end{figure}

Music audio and lyrics  provide complementary information in understanding
the richness of human beings' cultures and activities \cite{Nettl00}.
Music\footnote{https://en.wikipedia.org/wiki/Music} is an art expression
whose medium is sound organized in time. Lyrics\footnote{https://en.wikipedia.org/wiki/Lyrics}
as natural language represent music theme and story, which are a very
important element for creating a meaningful impression of the music.
Starting from the late 2014, Google provides music search results
containing song lyrics as shown in Fig.~\ref{fig:lyrics.pdf} when
given a specific song title. However, searching lyrics in this way
is insufficient because sometimes people might lack exact song title
but know a segment of music audio instead, or want to search an audio
track with part of the lyrics. Then, a natural question arises: how
to retrieve the lyrics by a segment of music audio, and vice versa?

Searching lyrics by audio was almost impossible years ago due to the
limited availability of large volumes of music audio and lyrics. The
profusion of online music audio and lyrics from music sharing websites
such as YouTube, MetroLyrics, Azlyrics, and Genius shows the opportunity
to understand musical knowledge from content-based audio and lyrics
by leveraging large volumes of cross-modal music data aggregated in
Internet.

Motivated by the fact that audio content and lyrics are very fundamental
aspects for understanding what kind of cultures and activities a song
wants to convey to us, this research pays attentions to deep correlation
learning between audio and lyrics for cross-modal music retrieval
and considers two real-world tasks: using audio to retrieve lyrics
or using lyrics to retrieve audio. Several contributions are made
in this paper, as follows:

i) To the best of our knowledge, this work is the first research where
a deep correlation learning architecture with two-branch neural networks
and correlation learning model is studied for cross-modal music retrieval
by using either audio or lyrics as a query modality.

ii) Different music modality data are projected to the shared space
where inter modal canonical correlation analysis is exploited as an
objective function to calculate the similarity of temporal structures.
Fully-connected deep neural networks (DNNs) and an end-to-end DNN
are proposed to learn audio representation, where the pre-trained
Doc2vec model followed by fully-connected layers is employed to extract
lyrics feature.

iii) Extensive experiments confirm the effectiveness of our deep correlation
learning architecture for audio-lyrics music retrieval, which are
meaningful results and studies for attracting more efforts on mining
music knowledge structure and correlation between different modality
data.

The rest of this paper is structured as follows. Research motivation
and background are introduced in Sec.\ref{sec:Related}. Sec.\ref{sec:preliminaries}
gives the preliminaries of Convolutional Neural Networks (CNNs) and
Deep Canonical Correlation Analysis (DCCA). Then, Sec.\ref{sec:Algorithm}
presents why and how we exploit CNNs and DCCA to build a deep correlation
learning architecture for audio-lyrics music retrieval. The task of
cross-modal music retrieval in our work is described in Sec.\ref{sec:Tasks}.
Experimental evaluation results are shown in Sec.\ref{sec:Experiments}.
Finally, conclusions are pointed out in Sec.\ref{sec:Conclusion}.

\section{Motivation and Background}

\label{sec:Related}Music has permeated our daily life, which contains
different modalities in real-world scenarios such as temporal audio
signal, lyrics with meaningful sentences, high-level semantic tags,
and temporal visual content. However, correlation learning between
lyrics and audio for cross-modal music retrieval has not been sufficiently
studied. Previous works \cite{Yu2010,Yu2013,Yu2009} mainly focused
on content-based music retrieval with single modality. With the widespread
availability of large-scale multimodal music data, it brings us research
opportunity to tackle cross-modal music retrieval.

\subsection{Lyrics and Audio in Music}

Recent research has shown that lyrics, audio, or the combination of
audio and lyrics are mainly applied to semantic classification such
as emotion or genre in music. For example, authors in \cite{McVicar2011}
proposed an unsupervised learning method for mood recognition where
Canonical Correlation Analysis (CCA) was applied to identify correlations
between lyrics and audio, and the evaluation of mood classification
was done based on the valence-arousal space. An interesting corpus
with each song in the MIDI format and emotion annotation is introduced
in \cite{Mihalcea2012}. Coarse-grained classification for six emotions
is learned by support vector machines (SVM), and this work showed
that either textual feature or audio feature can be used for emotion
classification, and their joint use leads to a significant improvement.
Emotion lyrics datasets in English \cite{Malheiro2016} are annotated
with continuous arousal and valence values. Specific text emotion
attributes are considered to complement music emotion recognition.
Experiments on the regression and classification of music lyrics by
quadrant, arousal, and valence categories are performed. Application
of hierarchical attention network is proposed in \cite{Alexandros2017}
to handle genre classification of intact lyrics. This network is able
to pay attention to words, lines, and segments of the song lyrics,
where the importance of words, lines, and segments in layer structure
is learned. Distinct from intensive research on music classification
by using lyrics and audio, our work focuses on audio-lyrics cross-modal
music retrieval: using audio to retrieve lyrics or vice versa. This
is a very natural way for us to retrieve lyrics or audio on the Internet.
However, no much research has investigated this task.

\subsection{Cross-modal Music Retrieval}

Some existing researches on cross-modal music retrieval intensively
focus on investigating music and visual modalities \cite{Yu2012,Acar2014,Mayer11,Brochu2003,Shah2014,Gillet2007,Nanni2016,Wu2016}.
Similarity between audio features extracted from music and image features
extracted from the album covers are trained by a Java SOMToolbox framework
in \cite{Mayer11}. Then, according to this similarity, people can
organize a music collection and make use of album cover as visual
content to retrieve a song over multimodal music data. Based on multi-modal
mixture models, a statistical method to jointly modeling music, images,
and text \cite{Brochu2003} is used to support retrieval over a multimodal
dataset. To generate a soundtrack for the outdoor video, an effective
heuristic ranking method is suggested based on heterogeneous late
fusion by jointly considering venue categories, visual scene, and
user listening history \cite{Shah2014}. Confidence scores, produced
by SVM-hmm models constructed from geographic, visual, and audio features,
are combined to obtain different types of video characteristics. To
learn the semantic correlation between music and video, a novel approach
to selecting features and statistical novelty based on kernel methods
\cite{Gillet2007} is proposed for music segmentation. Co-occurring
changes in audio and video content of music videos can be detected,
where the correlations can be used in cross-modal audio-visual music
retrieval. Lyrics-based music attributes are utilized for image representation
in \cite{Wu2016}. Cross-modal ranking analysis is suggested to learn
semantic similarity between music and image, with the aim of obtaining
the optimal embedding spaces for music and image. Distinct from intensive
research on considering the use of metadata for different music modalities
in cross-modal music retrieval, our work focuses on deep architecture
based on correlation learning between audio and lyrics for content-based
cross-modal music retrieval.

\subsection{Deep Cross-modal Learning}

We have witnessed several efforts devoted to investigating cross-modal
learning between different modalities, such as \cite{JiangL16,CaoL0L17,Yu2017,Zhong2017,HuangWW16,YuKCK16},
to facilitate cross-modal matching and retrieval. Most importantly,
latest studies extensively pay attention to deep cross-modal learning
between image and textual descriptions such as \cite{JiangL16,Yu2017,YuKCK16,YanM15}.
Most existing deep models with two-branch sub-networks explore pre-trained
convolutional neural network (CNN) \cite{Simonyan14} as image branch
\cite{Yu2017} and utilize pre-trained document-level embedding model
\cite{Lau16} or hand-crafted feature extraction such as bag of words
\cite{JiangL16} as text branch. Image and text modalities are converted
to the joint embedding space calculating a single ranking loss function
by feed-forward way. Image-text benchmarks such as \cite{LinMBHPRDZ14,Rasiwasia10}
are applied to evaluate the performances of cross-modal matching and
retrieval. There are two features for existing deep cross-modal retrieval:
i) cross-modal correlation between image and text is learned without
considering temporal sequences. ii) Pre-trained models are directly
applied to represent image or text. Distinct from existing deep cross-modal
retrieval architectures, this work takes into account temporal sequences
to learn the correlation between audio and lyrics for facilitating
audio-lyrics cross-modal music retrieval, where sequential audio and
lyrics are converted to the canonical space. A neural network with
two-branch sequential structures for audio and lyrics is trained.

\section{Preliminaries}

\label{sec:preliminaries}We focus on developing a two-branch deep
architecture for learning the correlation between audio and lyrics
in the cross-modal music retrieval, where several variants of deep
learning models are investigated for processing audio sequence while
pre-trained Doc2vec \cite{Lau16} is used for processing lyrics. A
brief review of CNNs and DCCA exploited in this work is addressed
in the following.

\subsection{Convolutional Neural Networks (CNNs)}

CNNs have been exploited to handle not only various tasks in the field
of computer vision and multimedia \cite{SimonyanZ2014,Krizhevsky2017},
but also the tasks of music information retrieval such as genre classification
\cite{Costa2017}, acoustic event detection \cite{Hershey2017}, automatic
music tagging \cite{Choi2016}. Generally speaking, when lacking computational
power and large annotated datasets, it is preferred to directly use
pre-trained CNNs such as VGG16 \cite{SimonyanZ2014} to extract features
\cite{Zhong2017}\cite{Hershey2017}, or \textcolor{black}{further
combine it with}\textcolor{red}{{} }fully-connected layers to extract
semantic features \cite{Yu2017}\cite{YanM15}\cite{Hershey2017}.

Different from plain spatial convolutional operation, CNN tries to
use different kernels (filters) to capture different local patterns,
and this will generate multiple intermediate feature maps (called
channels). Specifically, the convolutional operation in one convolutional
layer is defined as

\begin{equation}
\boldsymbol{x}^{j}=f(\sum_{k=0}^{K-1}\boldsymbol{H}^{jk}\otimes\boldsymbol{s}^{k}+a^{j}),\label{eq:conv}
\end{equation}
where the superscripts $j$, $k$ are channel indices, $\boldsymbol{s}^{k}$
is the $k$-th channel input, $\boldsymbol{x}^{j}$ is the $j$-th
channel output, $\otimes$ is the convolutional operation, $\boldsymbol{H}^{jk}$
is the convolutional kernel (or the filter) that associates the $k$-th
input channel with the $j$-th output channel, $a^{j}$ is the bias
for $j$-th channel, and $f(\cdot)$ is a non-linear activation function.
All weights that define a convolutional layer are represented as a
4-dimensional array with a shape of $(h,l,K,J)$, where $h$ and $l$
determine the kernel size, and $K$ and $J$ are the number of input
and output channels, respectively. When mel-spectrogram is used as
the input of the first convolutional layer, it only has one channel.

A 2D convolutional kernel $\boldsymbol{H}^{jk}$, as a common filter,
is applied to the whole input channel. This kernel is shifted along
both (frequency and time) axes and a local correlation is computed
between the kernel and input. The kernels are trained to find local
salient patterns that maximize the overall objective. As a kernel
sweeps the input, it generates a new output in order, which preserves
the spatiality of the input, i.e., the frequency and time constraint
of the spectrogram.

Convolutional layers are often followed by pooling layers, which reduce
the size of feature map by down sampling them. The max function is
a typical pooling operation. This selects the maximal value from a
pooling region, instead of keeping all information in the region.
This pooling operation also enables distortion and translation invariances
by discarding the original location of the selected value, and the
capability of such invariance within each pooling layer is determined
by the pooling size. With a small pooing size, the network does not
have enough distortion invariance, while a too large pooling size
may completely loose the location of a salient feature. Instead of
using a large pooling size in one layer, using multiple small pooling
sizes at different pooling layers will enable the system to gradually
abstract the features to be more compact and more semantic.

\subsection{Deep Canonical Correlation Analysis (DCCA)}

\label{sec:dcca-basic}CCA has been a very popular method for embedding
multimodal data in a shared space. Before presenting our deep multimodal
correlation learning between audio and lyrics, we first give an overview
of CCA and DCCA.

Let $\boldsymbol{x}\in R^{m}$ (e.g., audio feature) and $\boldsymbol{y}\in R^{n}$
(e.g., textual feature) be zero mean random (column) vectors with
covariances $\boldsymbol{C}_{xx}$, $\boldsymbol{C}_{yy}$ and cross-covariance
$\boldsymbol{C}_{xy}$. When a linear projection is performed, CCA
\cite{Hotelling36} tries to find two canonical weights $\boldsymbol{w}_{x}$
and $\boldsymbol{w}_{y}$, so that the correlation between the linear
projections $u=\boldsymbol{w}_{x}^{T}\boldsymbol{x}$ and $v=\boldsymbol{w}_{y}^{T}\boldsymbol{y}$
is maximized.

\begin{eqnarray}
(\boldsymbol{w}_{x},\boldsymbol{w}_{y}) & = & \underset{(\boldsymbol{w}_{x},\boldsymbol{w}_{y})}{argmax}\:corr(\boldsymbol{w}_{x}^{T}\boldsymbol{x},\boldsymbol{w}_{y}^{T}\boldsymbol{y})\nonumber \\
 & = & \underset{(\boldsymbol{w}_{x},\boldsymbol{w}_{y})}{argmax}\frac{{\boldsymbol{w}_{x}^{T}\boldsymbol{C}_{xy}\boldsymbol{w}_{y}}}{\sqrt{\boldsymbol{w}_{x}^{T}\boldsymbol{C}_{xx}\boldsymbol{w}_{x}\cdot\boldsymbol{w}_{y}^{T}\boldsymbol{C}_{yy}\boldsymbol{w}_{y}}}.\label{eq:CCA-vec}
\end{eqnarray}
One of the known shortcoming of CCA is that its linear projection
may not well model the nonlinear relation between different modalities.

DCCA \cite{Andrew13} tries to calculate non-linear correlations between
different modalities by a combination of DNNs (deep neural networks)
and CCA. Different from KCCA which relies on kernel functions (corresponding
to a logical high dimensional (sparse) space), DNN has the extra capability
of compressing features to a low dimensional (dense) space, and then
CCA is implemented in the objective function. The DNNs, which realize
the non-linear mapping ($\varphi_{x}(\cdot)$ and $\varphi_{y}(\cdot)$),
and the canonical weights ($\boldsymbol{w}_{x}$ and $\boldsymbol{w}_{y}$
that model the CCA between $\varphi_{x}(\boldsymbol{x})$ and $\varphi_{y}(\boldsymbol{y})$),
are trained simultaneously to maximize the correlation after the non-linear
mapping, as follows.

\begin{equation}
(\boldsymbol{w}_{x},\boldsymbol{w}_{y},\varphi_{x},\varphi_{y})=\underset{(\boldsymbol{w}_{x},\boldsymbol{w}_{y},\varphi_{x},\varphi_{y})}{argmax}\:corr(\boldsymbol{w}_{x}^{T}\varphi_{x}(\boldsymbol{x}),\boldsymbol{w}_{y}^{T}\varphi_{y}(\boldsymbol{y})).
\end{equation}

\section{Deep Audio-lyrics Correlation Learning}

\label{sec:Algorithm} We develop a deep cross-modal correlation learning
architecture that predicts latent alignment between audio and lyrics,
which enables audio-to-lyrics or lyrics-to-audio music retrieval.
In this section, we explain how our deep architecture is learned.
Specifically, we investigate different deep network models for correlation
analysis and different deep learning methods for audio feature extraction.

\subsection{Learning Strategy}

On one hand, lyrics as natural language express semantic music theme
and story; on the other hand, music audio contains some properties
such as tonality and temporal over time and frequency. They are correlated
in the semantic sense. However, audio and lyrics belong to different
modality and cannot be compared directly. Therefore, we extract their
features separately, and then map them to the same semantic space
for a similarity comparison. Because linear mapping in CCA does not
work well, we design deep networks to realize non-linear mapping before
CCA. Consequently, deep correlation models for learning temporal structures
are considered for representing lyrics branch and audio branch.

We investigate two deep network architectures. i) Separate feature
extraction, completely independent of the following DCCA analysis.
Text branch follows this architecture, where the pre-trained Doc2vec
\cite{Lau16} model is used to compute a compact textual feature vector.
 As for audio, directly using the pre-trained CNN model \cite{Choi2016}
belongs to this architecture as well. ii) Joint training of audio
feature extraction and DCCA analysis between audio and lyrics. In
this way, feature extraction is also correlated with the subsequent
DCCA. Here, for the audio branch, a CNN model is trained from the
ground together with the following fully-connected layers, based on
an end-to-end learning procedure. It is expected that this CNN is
adapted to the DNN so as to extract more meaningful audio features.

\subsection{Network Architecture}

\begin{figure}
\centering

\includegraphics[width=8cm]{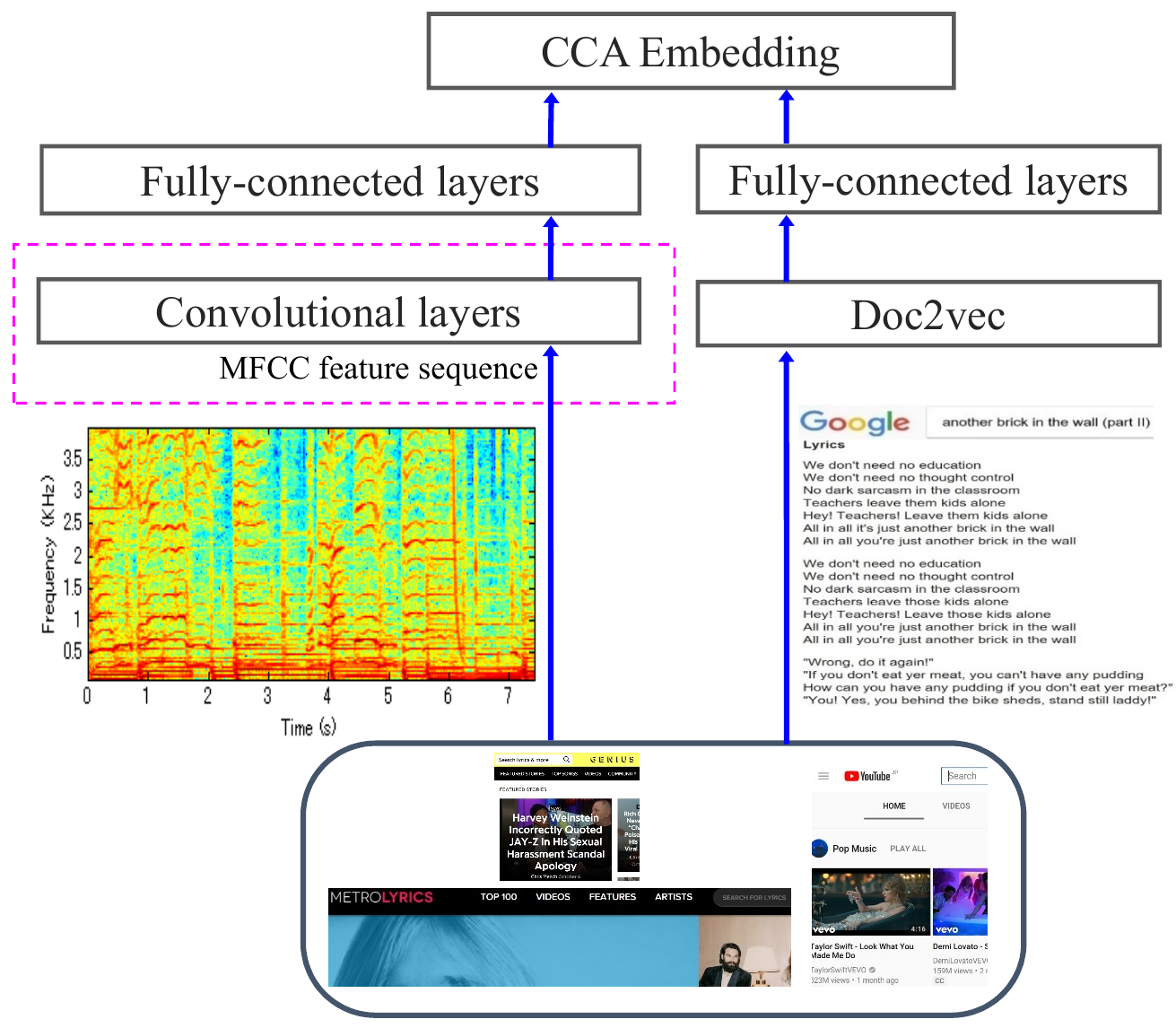}\caption{\label{fig:framework} Deep correlation learning between audio and
lyrics.}
\end{figure}

Figure~\ref{fig:framework} shows an end-to-end deep convolutional
DCCA network, which aims at simultaneously learning the feature extraction
and the deep correlation between audio and lyrics. This model is degenerated
to a simple DCCA network, when the CNN model marked in pink dashed
line is replaced by a pre-trained model.

From the sequence of words in the lyrics, textual feature is computed,
more specifically, by a pre-trained Doc2vec model. Music audio signal
is represented as a 2D spectrogram, which preserves both its spectral
and temporal properties. However, it is difficult to directly use
this for the DCCA analysis, due to its high dimension. Therefore,
we investigate two variants for the dimension reduction. (i) Audio
feature is extracted by a pre-trained convolutional model, and we
study the pure effect of DCCA in analyzing the correlation. i.e.,
sub DNNs with fully connected layers are trained to maximize the correlation
between audio and textual features. (ii) An end-to-end deep network
for audio branch that integrates convolutional layers for feature
extraction and non-linear mapping for correlation learning together,
is trained. In the future work, we will also consider the integration
of Doc2Vec with its subsequent DNN.

\subsubsection{Audio feature extraction}

\label{sec:4.2.1}The audio signal is represented as a spectrogram.
We mainly focus on mel-frequency cepstral coefficients (MFCCs), because
MFCCs are very efficient features for semantic genre classification
\cite{Sigtia2014} and music audio similarity comparison \cite{Hamel2013}.
We will also compare MFCC with Mel-spectrum, which contains more detailed
information. To compute a single feature vector for correlation analysis,
we successively apply convolutional layers with different kernels
to capture local salient features, and use pooling layers to reduce
the dimension.

By inserting the pooling layer between adjacent convolutional layers,
a kernel in the late layer corresponds to a larger kernel in the previous
layer, and has more capacity in representing semantic information.
Then, using small kernels in different convolutional layers can achieve
the function of a large kernel in one convolutional layer, but is
more robust to scale variance. In this sense, a combination of successive
convolutional layers and pooling layers can capture features at different
scales, and the kernels can learn to represent complex patterns.

For each audio signal, a slice of 30s is resampled to 22,050Hz with
a single channel. With frame length 2048 and step 1024, there are
646 frames. For the end-to-end learning, a sequence of MFCCs (20x646)
are computed. By initial experiments we found that our approach is
not very sensitive to the time resolution. Therefore, we decimate
the spectrogram into 4 sub sequences, each with 161 frames and associated
with the same lyrics.

For implementing an end-to-end deep learning, the configuration of
CNN used for audio branch in this work is shown in Table~\ref{tab:cnn-mfcc}.
It consists of 3 convolutional layers and 3 max pooling layers, and
outputs a feature vector with a size of 1536. We tried to add more
convolutional layer but see no significant difference. Rectified linear
unit (ReLU) is used as an activation function in each convolutional
layer except the last one. Batch normalization is used before activation.
Convolutional kernels (3x3) are used in every convolutional layer.
These kernels help to learn local spectral-tempo structures. In this
way, CNN converts an audio feature sequence (a 2D matrix) to a high
dimensional vector, and retains some astonishing properties such as
tempo invariances, which can be very helpful for learning musical
features in semantic correlation learning between lyrics and audio.

With the input spectrogram $\boldsymbol{s}$, the feature output by
the convolutional layers is $\boldsymbol{x}=f_{3}(\boldsymbol{H}_{3}\otimes f_{2}(\boldsymbol{H}_{2}\otimes f_{1}(\boldsymbol{H}_{1}\otimes\boldsymbol{s}+a_{1})+a_{2})+a_{3})$,
where $\boldsymbol{H}_{i}$, $a_{i}$ and $f_{i}$ are the convolutional
kernel, bias, and activation function in the $i$th layer.

\begin{table}[!t]
\caption{Configuration of CNNs for audio branch}

\label{tab:cnn-mfcc} 

\begin{centering}
\begin{tabular}{c}
\hline
MFCC: 20x646/4\tabularnewline
Convolution, 3x3x48\tabularnewline
Max-pooling (2,2), output 10x80x48\tabularnewline
Convolution: 3x3x96\tabularnewline
Max-pooling (3,3), output 3x26x96\tabularnewline
Convolution: 3x3x192\tabularnewline
Max-pooling (3,3), output 1536\tabularnewline
\hline
\end{tabular}
\par\end{centering}

\end{table}

As for the pre-trained model, we apply the pre-trained CNN model in
\cite{Choi2016}, which has 5 convolutional layers, each with either
average pooling or standard deviation pooling, generating a 30-dimension
vector per layer. Concatenating all of them together generates a feature
vector of 320 dimension.

\subsubsection{Textual feature extraction}

Lyrics text of each song is tokenized by using coreNLP \cite{ManningSBFBM14},
and passed to the infer\_vector module of the Doc2Vec model \cite{Lau16},
generating a 300-dimensional feature for each song. We use the pretrained
apnews\_dbow weights\footnote{https://ibm.ent.box.com/s/9ebs3c759qqo1d8i7ed323i6shv2js7e}
in the experiment.

\subsubsection{Non-linear mapping of features}

\label{sec:C-DCCA}Audio features and textual features are further
converted into low dimensional features in a shared $D$-dimensional
semantic space by using different sub DNNs composed of fully connected
layers.

The details of sub DNNs are shown in Table~\ref{tab:sub-dnn}. These
two sub DNNs (each with 3 fully connected layers) implement the non-linear
mapping of DCCA. The audio feature generated by the feature extraction
part is denoted as $\boldsymbol{x}\in R^{m}$ ($m$ varies with each
method) and deep textual feature is denoted as $\boldsymbol{y}\in R^{300}$.
The overall functions of sub-DNNs are denoted as $\varphi_{x}(\boldsymbol{x})=g_{3}(\boldsymbol{\Psi}_{3}\cdot g_{2}(\boldsymbol{\Psi}_{2}\cdot g_{1}(\boldsymbol{\Psi}_{1}\boldsymbol{x}+\boldsymbol{b}_{1})+\boldsymbol{b}_{2})+\boldsymbol{b}_{3})$,
where $\boldsymbol{\Psi}_{i}$ and $\boldsymbol{b}_{i}$ are the weight
matrix and bias for the $i$th layer and $g_{i}(\cdot)$ is the activation
function. And $\varphi_{y}(\boldsymbol{y})$ is computed in a similar
way. Then, $\varphi_{x}(\boldsymbol{x})$ is the overall result of
the convolutional layer and its subsequent DNN, given the input spectrogram
$\boldsymbol{s}$.

\begin{table}[!t]
\caption{Structure of sub-DNNs}

\label{tab:sub-dnn} 

\begin{centering}
\begin{tabular}{c|c|c}
\hline
 & Sub-DNN1 (Audio)  & Sub-DNN2 (Text)\tabularnewline
1st layer  & 1024, sigmoid  & 1024, sigmoid\tabularnewline
2nd layer  & 1024, sigmoid  & 1024, sigmoid\tabularnewline
3rd layer (output)  & $D$, linear  & $D$, linear\tabularnewline
\hline
\end{tabular}
\par\end{centering}

\end{table}

\subsubsection{Objective function of CCA}

Assume the batch size in the training is $N$, $\boldsymbol{X}\in R^{D\times N}$
and $\boldsymbol{Y}\in R^{D\times N}$ are the outputs of sub DNN
of the two batches, corresponding to audio ($\varphi_{x}(\boldsymbol{x})$)
and lyrics ($\varphi_{y}(\boldsymbol{y})$ ), respectively. Let covariance
of $\varphi_{x}(\boldsymbol{x})$ and $\varphi_{y}(\boldsymbol{y})$
be $\boldsymbol{C}_{XX}$, $\boldsymbol{C}_{YY}$ and their cross-covariance
be $\boldsymbol{C}_{XY}$. With the linear projection matrices $\boldsymbol{W}_{X}$
and $\boldsymbol{W}_{Y}$, the correlation between the canonical components
($\boldsymbol{W}_{X}^{T}\boldsymbol{X}$ and $\boldsymbol{W}_{Y}^{T}\boldsymbol{Y}$)
can be computed. This correlation indicates the association between
the two modalities and is used as an overall objective function, which
is maximized to find all parameters (convolutional kernels $\boldsymbol{H}(\cdot)$,
non-linear projections $\varphi_{x}(\cdot)$ and $\varphi_{y}(\cdot)$,
linear projection matrices $\boldsymbol{W}_{X}$ and $\boldsymbol{W}_{Y}$).

\begin{eqnarray*}
(\boldsymbol{H},\boldsymbol{W}_{X},\negthinspace\boldsymbol{W}_{Y},\negthinspace\varphi_{x},\negthinspace\varphi_{y}) & \negthinspace\negthinspace\negthinspace\negthinspace\negthinspace=\negthinspace\negthinspace\negthinspace\negthinspace\negthinspace\negthinspace\negthinspace\negthinspace\negthinspace\negthinspace & \underset{(\boldsymbol{H},\boldsymbol{W}_{X},\boldsymbol{W}_{Y},\varphi_{x},\varphi_{y})}{argmax}\negthinspace\negthinspace\negthinspace\negthinspace corr(\boldsymbol{W}_{X}^{T}\boldsymbol{X},\boldsymbol{W}_{Y}^{T}\boldsymbol{Y}).
\end{eqnarray*}
At first, with $\boldsymbol{H},\varphi_{x},\varphi_{y}$ being fixed,
$\boldsymbol{W}_{X}$ and $\boldsymbol{W}_{Y}$ are computed by

\begin{eqnarray*}
(\boldsymbol{W}_{X},\boldsymbol{W}_{Y}) & \negthinspace\negthinspace\negthinspace\negthinspace=\negthinspace\negthinspace\negthinspace\negthinspace & \underset{(\boldsymbol{W}_{X},\boldsymbol{W}_{Y})}{argmax}\frac{{\boldsymbol{W}_{X}^{T}\boldsymbol{C}_{XY}\boldsymbol{W}_{Y}}}{\sqrt{\boldsymbol{W}_{X}^{T}\boldsymbol{C}_{XX}\boldsymbol{W}_{X}\cdot\boldsymbol{W}_{Y}^{T}\boldsymbol{C}_{YY}\boldsymbol{W}_{Y}}}.
\end{eqnarray*}
This can be rewritten in the trace-form

\begin{eqnarray}
(\boldsymbol{W}_{X},\negthinspace\boldsymbol{W}_{Y}) & \negthinspace\negthinspace\negthinspace\negthinspace\negthinspace=\negthinspace\negthinspace\negthinspace\negthinspace\negthinspace\negthinspace\negthinspace\negthinspace & \underset{(\boldsymbol{W}_{X},\boldsymbol{W}_{Y})}{argmax}tr(\boldsymbol{W}_{X}^{T}\boldsymbol{C}_{XY}\boldsymbol{W}_{Y}),\label{eq:corr-trace}\\
\text{{subject\,to}}: & \negthinspace\negthinspace\negthinspace\negthinspace\negthinspace\negthinspace\negthinspace & \negthinspace\negthinspace\negthinspace\negthinspace\boldsymbol{W}_{X}^{T}\boldsymbol{C}_{XX}\boldsymbol{W}_{X}\negthinspace=\negthinspace\boldsymbol{W}_{Y}^{T}\boldsymbol{C}_{YY}\boldsymbol{W}_{Y}\negthinspace=\negthinspace\boldsymbol{I}.\nonumber
\end{eqnarray}
Here, covariance $\boldsymbol{C}_{XX}$, $\boldsymbol{C}_{YY}$ and
cross-covariance $\boldsymbol{C}_{XY}$ are computed as follows

\begin{equation}
\boldsymbol{C}_{XX}=\frac{{1}}{N-1}\hat{\boldsymbol{X}}\hat{\boldsymbol{X}}^{T}+r\boldsymbol{I},
\end{equation}

\begin{equation}
\boldsymbol{C}_{YY}=\frac{{1}}{N-1}\hat{\boldsymbol{Y}}\hat{\boldsymbol{Y}}^{T}+r\boldsymbol{I},
\end{equation}

\begin{equation}
\boldsymbol{C}_{XY}=\frac{{1}}{N-1}\hat{\boldsymbol{X}}\hat{\boldsymbol{Y}}^{T},
\end{equation}

\[
\hat{\boldsymbol{X}}=\boldsymbol{X}-\boldsymbol{\overline{X}},\hat{\boldsymbol{Y}}=\boldsymbol{Y}-\boldsymbol{\overline{Y}}
\]
where $\boldsymbol{\overline{X}}$ and $\boldsymbol{\overline{Y}}$
are average of $\varphi_{x}(\boldsymbol{x})$ and $\varphi_{y}(\boldsymbol{y})$
within the batch, and $r$ is a small positive constant used to ensure
the positive definiteness of $\boldsymbol{C}_{XX}$ and $\boldsymbol{C}_{YY}$.

By defining $\boldsymbol{T}\triangleq\boldsymbol{C}_{XX}^{-1/2}\boldsymbol{C}_{XY}\boldsymbol{C}_{YY}^{-1/2}$
and performing singular value decomposition on $\boldsymbol{T}$ as
$\boldsymbol{\boldsymbol{T}=U}\boldsymbol{D}\boldsymbol{V}^{T}$,
$\boldsymbol{W}_{X}$ and $\boldsymbol{W_{Y}}$ can be computed by
\cite{Andrew13}

\begin{equation}
\boldsymbol{W}_{X}=\boldsymbol{C}_{XX}^{-1/2}\boldsymbol{U},\boldsymbol{W}_{Y}=\boldsymbol{C}_{YY}^{-1/2}\boldsymbol{V}.\label{eq:CCA-sol-1}
\end{equation}
Then, Eq.(\ref{eq:corr-trace}) can be rewritten as

\begin{equation}
tr((\boldsymbol{W}_{X}^{T}\boldsymbol{C}_{XY}\boldsymbol{W}_{Y})^{T}\cdot\boldsymbol{W}_{X}^{T}\boldsymbol{C}_{XY}\boldsymbol{W}_{Y})=tr(\boldsymbol{T}^{T}\boldsymbol{T}).
\end{equation}
Accordingly, the gradient of the correlation with respect to $\boldsymbol{X}$
is given by

\begin{equation}
\frac{1}{N-1}(2\nabla_{XX}\hat{\boldsymbol{X}}+\nabla_{XY}\hat{\boldsymbol{Y}}),
\end{equation}

\[
\nabla_{XX}=-\frac{1}{2}\boldsymbol{C}_{XX}^{-1/2}\boldsymbol{U}\boldsymbol{D}\boldsymbol{U}^{T}\boldsymbol{C}_{XX}^{-1/2},
\]

\[
\nabla_{XY}=\boldsymbol{C}_{XX}^{-1/2}\boldsymbol{U}\boldsymbol{V}^{T}\boldsymbol{C}_{YY}^{-1/2}.
\]
And the gradient of the correlation with respect to $\boldsymbol{Y}$
can be computed in a similar way.

Then, the gradients are back propagated, first in the sub DNN, where
$\varphi_{x}(\boldsymbol{x})$ and $\varphi_{x}(\boldsymbol{y})$
are updated. As for the audio branch, the gradients are further back
propagated to the convolutional layers, and the kernel filters $\boldsymbol{H}$
are updated. The whole procedure is shown in Algorithm~\ref{alg:JointTrain}.
\begin{algorithm}
\begin{algorithmic}[1]
\Procedure{JointTrain}{$\boldsymbol{A}, \boldsymbol{L}$}
\Comment{$\boldsymbol{A}$: audio, $\boldsymbol{L}$: lyrics} %
   \State Initialize convolutional net, sub-networks for mapping  %
   \State Compute MFCC spectrogram from audio $\boldsymbol{A}, \rightarrow \boldsymbol{\Omega}_A$ %
   \State Compute textual feature from lyrics $\boldsymbol{L}, \rightarrow \boldsymbol{\Omega}_L$  %
   \For{each epoch}
      \State Randomly divide $\boldsymbol{\Omega}_A, \boldsymbol{\Omega}_L$ to batches %
      \For{each batch ($\boldsymbol{\omega}_A, \boldsymbol{\omega}_L$) of audio and lyrics}
         \For{each pair $(\boldsymbol{s}, \boldsymbol{l})$ $\in (\boldsymbol{\omega}_A, \boldsymbol{\omega}_L)$}                       \State  $\boldsymbol{s} \rightarrow \boldsymbol{x}$ by convolutions %
            \State  $\boldsymbol{l} \rightarrow \boldsymbol{y}$ by pretrained Doc2Vec model %
            \State  $\boldsymbol{x} \rightarrow \varphi_{x}(\boldsymbol{x})$ by non-linear mapping %
            \State  $\boldsymbol{y} \rightarrow \varphi_{y}(\boldsymbol{y})$ by non-linear mapping %
         \EndFor
         \State  Get converted batch ($\boldsymbol{X}, \boldsymbol{Y})$
         \State  Apply CCA on ($\boldsymbol{X}, \boldsymbol{Y}$) to compute $\boldsymbol{W}_X, \boldsymbol{W}_Y$ %
         \State  Compute the gradient with respect to $\boldsymbol{X}, \boldsymbol{Y}$ %
         \State  Back propagate to the sub network %
         \State  Back propagate to the convolutional network %
      \EndFor
   \EndFor
\EndProcedure
\end{algorithmic}

\caption{\label{alg:JointTrain}Joint training of CNN and DCCA}

\end{algorithm}

\section{Music cross-modal retrieval tasks}

\label{sec:Tasks}Two kinds of retrieval tasks are defined to evaluate
the effectiveness of our algorithms: instance-level and category-level.
Instance-level cross-modal music retrieval is to retrieve lyrics when
given music audio as input or vice versa. Category-level cross-modal
music retrieval is to retrieve lyrics or audio, searching most similar
audio or lyrics with the same mood category.

With a given input (either audio slice or lyrics), its canonical component
is computed, and its similarity with the canonical components of the
other modality in the database is computed using the cosine similarity
metric, and the results are ranked in the decreasing order of the
similarity score.

\section{Experiments}

\label{sec:Experiments}  The performances of the proposed DCCA variants
are evaluated and compared with some baselines such as variants of
CCA and deep multi-view embedding approach \cite{HeWL16}.

\subsection{Experiment Setting}

\emph{Proposed methods}. As discussed in Sec.~\ref{sec:Algorithm},
two variants of DCCA in combination with CNN are investigated: 1)
PretrainCNN-DCCA (the application of DCCA on the pretrained CNN model
\cite{Choi2016}), 2) JointTrain-DCCA (the joint training of CNN and
DCCA).

\emph{Baseline methods} include some shallow correlation learning
methods (without fully connected layers between feature extraction
and CCA), such as 3) Spotify-CCA (which applies CCA on the 65-dimensional
audio features provided by Spotify\footnote{https://developer.spotify.com/web-api/get-audio-features/}),
4) PretrainCNN-CCA (which applies CCA on the features extracted by
the pretrained CNN model), and multi-view methods such as 5) Spotify-MVE
(Spotify feature with deep multi-view embedding method similar to
\cite{HeWL16} where arbitrary mappings of two different views are
embedded in the joint space based on considering matched pairs with
minimal distance and mismatched pairs with maximal distance), 6) PretrainCNN-MVE.
We also evaluated 7) Spotify-DCCA. In all these methods, the lyrics
branch uses the features extracted by the pretrained Doc2vec model.

Besides MFCC, we also evaluate the feature of Mel-spectrum. The dimension
for Mel-spectrum is 96 per frame, and there are four convolutional
layers, where each of the first three is followed by a max pooling
layer, and the final output is 3072 dimension. As for the MVE methods,
both branches share the same parameters (activation function, number
of neurons and so on) and both have 3 fully connected layers (with
512, 256, and 128 neurons respectively). Batch normalization is used
before each layer and tanh activation function is applied after each
layer.

\emph{Audio-lyrics dataset}. Currently, there is no large audio/lyrics
dataset publically available for cross-modal music retrieval. Therefore,
we build a new audio-lyrics dataset. Spotify is a music streaming
on-demand service, which provides access to over 30 million songs,
where songs can be searched by various parameters such as artist,
playlist, and genre. Users can create, edit, and share playlists on
Spotify. Initially, we take 20 most frequent mood categories (aggressive,
angry, bittersweet, calm, depressing, dreamy, fun, gay, happy, heavy,
intense, melancholy, playful, quiet, quirky, sad, sentimental, sleepy,
soothing, sweet) \cite{Yu2012} as playlist seeds to invoke Spotify
API. For each mood category, we find the top 500 popular English songs
according to the popularity provided by Spotify, and further crawl
30s audio slices of these songs from YouTube, while lyrics are collected
from Musixmatch. Altogether there are 10,000 pairs of audio and lyrics.

\emph{Evaluation metric}. In the retrieval evaluation, we use mean
reciprocal rank 1 (MRR1) and recall@N as the metrics. Because there
is only one relevant audio or lyrics, MRR1 is able to show the rank
of the result. MRR1 is defined by

\begin{equation}
MRR1=\frac{1}{N_{q}}\sum_{i=1}^{N_{q}}\frac{1}{rank_{i}(1)},
\end{equation}
where $N_{q}$ is the number of the queries and $rank_{i}(1)$ corresponds
to the rank of the relevant item in the $i$th query. We also evaluate
recall@N to see how often the relevant item is included in the top
of the ranked list. Assume $S_{q}$ is the set of its relevant items
($|S_{q}|=1$) in the database for a given query and the system outputs
a ranked list $K_{q}$ ($|K_{q}|=N$). Then, recall is computed by

\begin{equation}
recall=\frac{|S_{q}\text{\ensuremath{\bigcap}}K_{q}|}{|S_{q}|}
\end{equation}
and is averaged over all queries.

We use 8,000 pairs of audio and lyrics as the training dataset, and
the rest 2,000 pairs for the retrieval testing. Because we generate
4 sub-sequences from each original MFCC sequence, there are 32,000
pairs of audio/lyric pairs in JoinTrain. In each run, the split of
audio-lyrics pairs into training/testing is random, and a new model
is trained. All results are averaged over 5 runs (cross-validations).
In the batch-based training, the batch size is unified to 1000 samples
in all methods, and the training takes 200 epochs for JointTrain and
400 epochs for other DCCA methods. Furthermore, training MVE requires
the presence of non-paired instances. To this end, we randomly selected
1 non-paired instance for each song in the dataset. The margin hyper-parameter
was set to 0.3, according to our preliminary experiments. Then, we
trained MVE for 1280 epochs.

\emph{Experiment environment}. The evaluations are performed on a
Centos7.2 server, which is configured with two E5-2620v4 CPU (2.1GHz),
three GTX 1080 GPU (11GB), and DDR4-2400 Memory (128G). Moreover,
it contains CUDA8.0, Conda3-4.3 (python 3.5), Tensorflow 1.3.0, and
Keras 2.0.5.

\subsection{Performance under Different Numbers of Epochs}

\begin{figure}[tb]
\centering

\includegraphics[width=8cm]{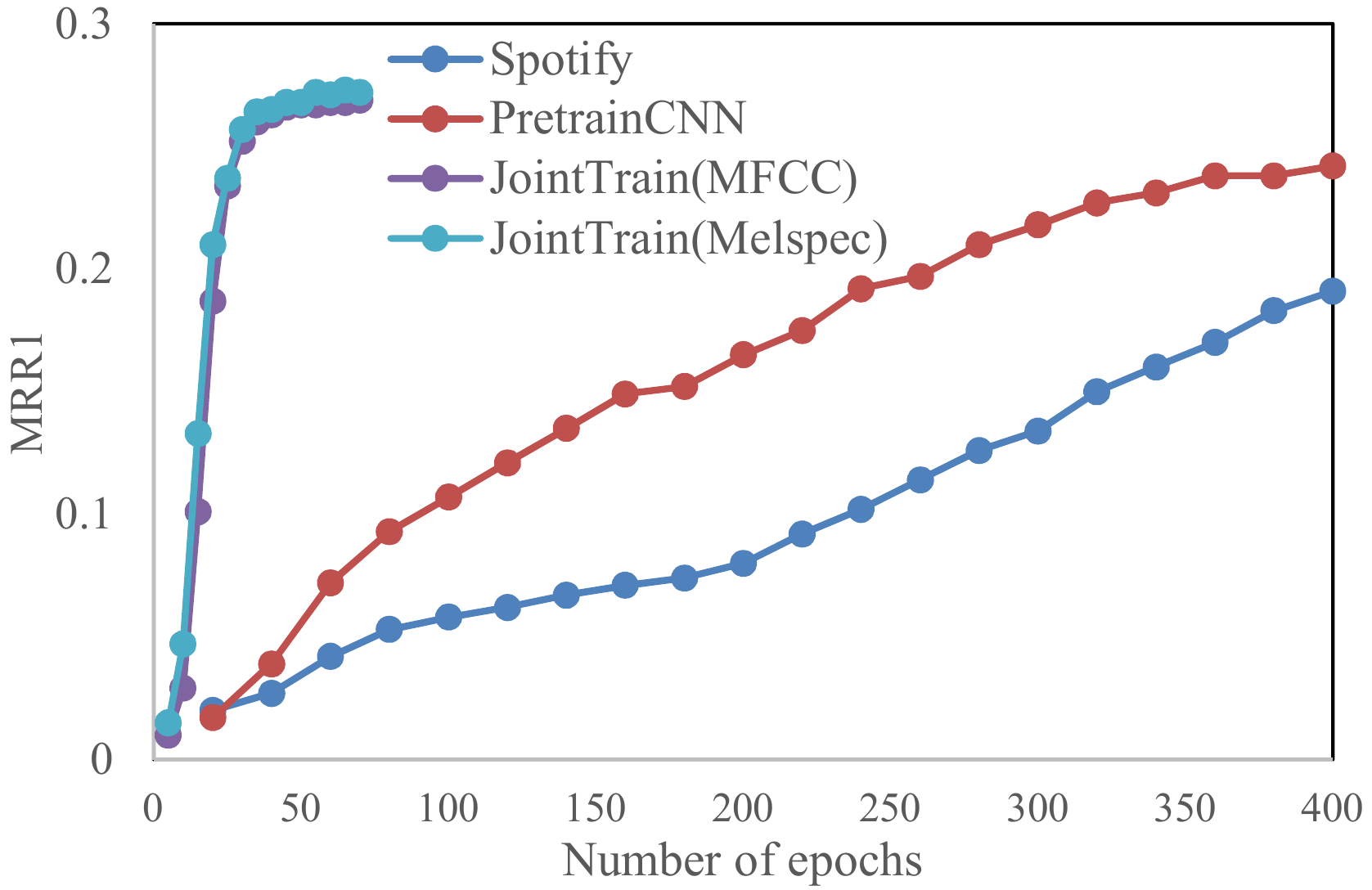}\caption{\label{fig:mrr1a_iter} MRR1 with respect to the numbers of epochs
(Using audio as query to search lyrics, \#CCA-component=30)}
\end{figure}

Fig.~\ref{fig:mrr1a_iter} shows the MRR1 results of Spotify-DCCA,
PretrainCNN-DCCA, JointTrain-DCCA with MFCC and JointTrain-DCCA with
Mel-spectrum, under different numbers of epochs. In all methods, MRR1
increases with the number of epochs, but with different trend. It
is clear that MFCC has similar performance as Mel-spectrum, converging
much fast than the other two methods and achieving higher MRR1. Hereafter,
we only use MFCC as the raw feature for JointTrain.

\begin{table*}[!t]
\caption{Instance-level MRR1 with respect to different numbers of CCA/MVE components
(Using audio as query)}

\label{tab:mrr1a_dim} 

\begin{centering}
\begin{tabular}{c|c|c|c|c|c|c|c}
\hline
\#CCA/MVE & Spotify-CCA & PretrainCNN-CCA & Spotify-MVE & PretrainCNN-MVE & Spotify-DCCA & PretrainCNN-DCCA & JointTrain-DCCA\tabularnewline
\hline
10 & 0.023 & 0.022 & 0.121 & 0.166 & 0.125 & 0.189 & 0.247\tabularnewline
20 & 0.029 & 0.040 & 0.134 & 0.187 & 0.168 & 0.225 & 0.254\tabularnewline
30 & 0.034 & 0.054 & 0.095 & 0.158 & 0.183 & 0.236 & 0.256\tabularnewline
40 & 0.039 & 0.069 & 0.084 & 0.115 & 0.183 & 0.239 & 0.256\tabularnewline
50 & 0.039 & 0.078 & 0.067 & 0.107 & 0.178 & 0.237 & 0.256\tabularnewline
60 & 0.040 & 0.085 & 0.065 & 0.094 & 0.177 & 0.240 & \textcolor{black}{0.257}\tabularnewline
70 & N/A & 0.090 & 0.061 & 0.085 & 0.174 & 0.239 & 0.256\tabularnewline
80 & N/A & 0.094 & 0.056 & 0.080 & 0.171 & 0.237 & 0.257\tabularnewline
90 & N/A & 0.098 & 0.054 & 0.063 & 0.164 & 0.238 & 0.257\tabularnewline
100 & N/A & 0.099 & 0.043 & 0.072 & 0.154 & 0.237 & 0.257\tabularnewline
\hline
\end{tabular}
\par\end{centering}

\end{table*}

\subsection{Impact of the Numbers of CCA Components}

Here, we evaluate the impact of the number of CCA/MVE components,
which affects the performance of both the baseline methods and the
proposed methods. The number of CCA/MVE components is adjusted from
10 to 100. The results of MRR1 and recall of Spotify-CCA are marked
as N/A when the number of CCA components is greater than 65, the dimension
of Spotify feature.

The MRR1 results, with audio feature as query to search lyrics, are
shown in Table~\ref{tab:mrr1a_dim}. Clearly, with the linear CCA,
Spotify-CCA and PretrainCNN-CCA have poor performance, although the
performance increases with the number of CCA components. In comparison,
with DCCA, the MRR1 results are much improved in Spotify-DCCA and
PretrainCNN-DCCA. The MRR1 performance increases with the number of
CCA components, and approaches a constant value in PretrainCNN-DCCA.
MRR1 decreases a little in Spotify-DCCA when the number of CCA components
gets greater than 65, the dimension of Spotify feature. Using MVE,
the peak performance of Spotify-MVE and PretrainCNN-MVE lies between
that of CCA and DCCA. With the end-to-end training, the MRR1 performance
is further improved in JointTrain-DCCA, and is almost insensitive
to the number of CCA components. But a further increase in the number
of CCA components will lead to the SVD failure in CCA.

Table~\ref{tab:mrr1t_dim} shows the MRR1 results achieved using
lyrics as query to search audio in the database, which has a similar
trend as in Table~\ref{tab:mrr1a_dim}. Generally, when audio and
lyrics are converted to the same semantic space, they share the same
statistics, and can be retrieved mutually.

\begin{table*}[!t]
\caption{Instance-level MRR1 with respect to different numbers of CCA/MVE components
(Using lyrics as query)}

\label{tab:mrr1t_dim} 

\begin{centering}
\begin{tabular}{c|c|c|c|c|c|c|c}
\hline
\#CCA/MVE & Spotify-CCA & PretrainCNN-CCA & Spotify-MVE & PretrainCNN-MVE & Spotify-DCCA & PretrainCNN-DCCA & JointTrain-DCCA\tabularnewline
\hline
10 & 0.022 & 0.022 & 0.114 & 0.157 & 0.124 & 0.190 & 0.248\tabularnewline
20 & 0.029 & 0.038 & 0.119 & 0.179 & 0.168 & 0.225 & 0.254\tabularnewline
30 & 0.034 & 0.053 & 0.083 & 0.147 & 0.184 & 0.236 & 0.256\tabularnewline
40 & 0.038 & 0.065 & 0.067 & 0.100 & 0.183 & 0.240 & 0.254\tabularnewline
50 & 0.041 & 0.076 & 0.056 & 0.097 & 0.180 & 0.236 & 0.256\tabularnewline
60 & 0.041 & 0.083 & 0.053 & 0.082 & 0.176 & \textcolor{black}{0.241} & \textcolor{black}{0.257}\tabularnewline
70 & N/A & 0.089 & 0.049 & 0.074 & 0.174 & 0.240 & 0.256\tabularnewline
80 & N/A & 0.094 & 0.048 & 0.068 & 0.170 & 0.237 & 0.257\tabularnewline
90 & N/A & 0.099 & 0.044 & 0.053 & 0.163 & 0.239 & 0.256\tabularnewline
100 & N/A & 0.102 & 0.035 & 0.062 & 0.152 & 0.237 & 0.256\tabularnewline
\hline
\end{tabular}
\par\end{centering}

\end{table*}

Table~\ref{tab:recalla_dim} and Table~\ref{tab:recallt_dim} show
the results of recall@1 and result@5. Recall@N in these tables is
only a little greater than MRR1 in Table~\ref{tab:mrr1a_dim} and
Table~\ref{tab:mrr1t_dim}, which indicates that for most queries,
its relevant item either appears at the first place, or not in the
top-n list at all. This infers that for some songs, lyrics and audio,
even after being mapped to the same semantic space, are not similar
enough.

\begin{table*}[!t]
\caption{Instance-level Recall @ N with respect to different numbers of CCA
components (Using audio as query)}

\label{tab:recalla_dim} 

\begin{centering}
\begin{tabular}{c|cc|cc|c|cc|cc|c}
\hline
 & \multicolumn{2}{c|}{Spotify@1} & \multicolumn{2}{c|}{PretrainCNN@1} & \multicolumn{1}{c|}{JointTrain@1} & \multicolumn{2}{c|}{Spotify@5} & \multicolumn{2}{c|}{PretrainCNN@5} & JointTrain@5\tabularnewline
\hline
 & CCA & DCCA & CCA & DCCA & DCCA & CCA & DCCA & CCA & DCCA & DCCA\tabularnewline
\hline
10 & 0.006 & 0.094 & 0.007 & 0.160 & 0.233 & 0.025 & 0.150 & 0.025 & 0.217 & 0.257\tabularnewline
20 & 0.010 & 0.138 & 0.020 & 0.204 & 0.243 & 0.034 & 0.193 & 0.047 & 0.243 & 0.262\tabularnewline
30 & 0.014 & 0.155 & 0.031 & 0.217 & 0.245 & 0.043 & 0.205 & 0.068 & 0.252 & 0.263\tabularnewline
40 & 0.019 & 0.155 & 0.045 & \textcolor{black}{0.221} & \textcolor{black}{0.245} & \textcolor{black}{0.047} & \textcolor{black}{0.205} & \textcolor{black}{0.085} & \textcolor{black}{0.255} & \textcolor{black}{0.262}\tabularnewline
50 & 0.020 & 0.150 & 0.053 & \textcolor{black}{0.220} & \textcolor{black}{0.246} & \textcolor{black}{0.049} & \textcolor{black}{0.200} & \textcolor{black}{0.095} & \textcolor{black}{0.250} & \textcolor{black}{0.262}\tabularnewline
60 & 0.020 & 0.151 & 0.060 & \textcolor{black}{0.222} & \textcolor{black}{0.246} & \textcolor{black}{0.051} & \textcolor{black}{0.197} & \textcolor{black}{0.102} & \textcolor{black}{0.254} & \textcolor{black}{0.263}\tabularnewline
70 & N/A & 0.147 & 0.065 & \textcolor{black}{0.222} & \textcolor{black}{0.246} & N/A & \textcolor{black}{0.197} & \textcolor{black}{0.107} & \textcolor{black}{0.253} & \textcolor{black}{0.263}\tabularnewline
80 & N/A & 0.144 & 0.068 & \textcolor{black}{0.220} & \textcolor{black}{0.246} & N/A & \textcolor{black}{0.191} & \textcolor{black}{0.112} & \textcolor{black}{0.250} & \textcolor{black}{0.264}\tabularnewline
90 & N/A & 0.137 & 0.071 & \textcolor{black}{0.220} & \textcolor{black}{0.247} & N/A & \textcolor{black}{0.186} & \textcolor{black}{0.120} & \textcolor{black}{0.253} & \textcolor{black}{0.263}\tabularnewline
100 & N/A & 0.129 & 0.073 & 0.220 & 0.246 & N/A & 0.175 & 0.121 & 0.251 & 0.263\tabularnewline
\hline
\end{tabular}
\par\end{centering}

\end{table*}

\begin{table*}[!t]
\caption{Instance-level Recall @ N with respect to different numbers of CCA
components (Using lyrics as query)}

\label{tab:recallt_dim} 

\begin{centering}
\begin{tabular}{c|cc|cc|c|cc|cc|c}
\hline
 & \multicolumn{2}{c|}{Spotify@1} & \multicolumn{2}{c|}{PretrainCNN@1} & \multicolumn{1}{c|}{JointTrain@1} & \multicolumn{2}{c|}{Spotify@5} & \multicolumn{2}{c|}{PretrainCNN@5} & JointTrain@5\tabularnewline
\hline
 & CCA & DCCA & CCA & DCCA & DCCA & CCA & DCCA & CCA & DCCA & DCCA\tabularnewline
\hline
10 & 0.005 & 0.090 & 0.007 & 0.160 & 0.235 & 0.024 & 0.151 & 0.022 & 0.219 & 0.257\tabularnewline
20 & 0.009 & 0.138 & 0.019 & 0.204 & 0.242 & 0.034 & 0.193 & 0.048 & 0.242 & 0.261\tabularnewline
30 & 0.014 & 0.157 & 0.031 & 0.219 & 0.245 & 0.042 & 0.205 & 0.064 & 0.250 & 0.263\tabularnewline
40 & 0.018 & 0.155 & 0.040 & 0.223 & 0.244 & 0.048 & 0.205 & 0.081 & 0.252 & 0.261\tabularnewline
50 & 0.021 & 0.154 & 0.050 & 0.218 & 0.246 & 0.051 & 0.199 & 0.092 & 0.250 & 0.262\tabularnewline
60 & 0.021 & 0.150 & 0.057 & \textcolor{black}{0.224} & \textcolor{black}{0.247} & \textcolor{black}{0.051} & \textcolor{black}{0.197} & \textcolor{black}{0.101} & \textcolor{black}{0.254} & \textcolor{black}{0.263}\tabularnewline
70 & N/A & 0.147 & 0.064 & \textcolor{black}{0.224} & \textcolor{black}{0.245} & N/A & \textcolor{black}{0.196} & \textcolor{black}{0.108} & \textcolor{black}{0.252} & \textcolor{black}{0.263}\tabularnewline
80 & N/A & 0.144 & 0.069 & \textcolor{black}{0.221} & \textcolor{black}{0.247} & N/A & \textcolor{black}{0.190} & \textcolor{black}{0.113} & \textcolor{black}{0.250} & \textcolor{black}{0.264}\tabularnewline
90 & N/A & 0.137 & 0.072 & 0.222 & 0.246 & N/A & 0.186 & 0.119 & 0.253 & 0.263\tabularnewline
100 & N/A & 0.126 & 0.077 & 0.221 & 0.247 & N/A & 0.172 & 0.121 & 0.249 & 0.262\tabularnewline
\hline
\end{tabular}
\par\end{centering}

\end{table*}

Table~\ref{tab:mrr1aclass_dim} and Table~\ref{tab:mrr1tclass_dim}
show the MRR1 results per category, where the first item with the
same mood category as the query is regarded as relevant. Compared
with the instance-level retrieval, the MRR1 result per category is
about 12\% larger in all methods, but cannot be improved more by increasing
the number of CCA/MVE components. Because there are 20 mood categories,
and some mood categories have similar meaning, this increases the
difficulty of distinguishing songs in the category level.

\begin{table*}[!t]
\caption{Category-level MRR1 with respect to different numbers of CCA/MVE components
(Using audio as query)}

\label{tab:mrr1aclass_dim} 

\begin{centering}
\begin{tabular}{c|c|c|c|c|c|c|c}
\hline
\#CCA/MVE & Spotify-CCA & PretrainCNN-CCA & Spotify-MVE & PretrainCNN-MVE & Spotify-DCCA & Pretrain-DCCA & JointTrain-DCCA\tabularnewline
\hline
10 & 0.177 & 0.172 & 0.249 & 0.286 & 0.260 & 0.313 & 0.364\tabularnewline
20 & 0.180 & 0.187 & 0.265 & 0.313 & 0.296 & 0.344 & 0.367\tabularnewline
30 & 0.182 & 0.199 & 0.230 & 0.284 & 0.307 & 0.349 & 0.372\tabularnewline
40 & 0.187 & 0.212 & 0.222 & 0.246 & 0.307 & \textcolor{black}{0.356} & \textcolor{black}{0.368}\tabularnewline
50 & 0.189 & 0.218 & 0.211 & 0.237 & 0.304 & \textcolor{black}{0.358} & \textcolor{black}{0.370}\tabularnewline
60 & 0.188 & 0.225 & 0.206 & 0.230 & 0.302 & \textcolor{black}{0.355} & \textcolor{black}{0.373}\tabularnewline
70 & N/A & 0.230 & 0.203 & 0.221 & 0.298 & \textcolor{black}{0.358} & \textcolor{black}{0.370}\tabularnewline
80 & N/A & 0.234 & 0.196 & 0.215 & 0.294 & \textcolor{black}{0.352} & \textcolor{black}{0.370}\tabularnewline
90 & N/A & 0.235 & 0.192 & 0.203 & 0.294 & \textcolor{black}{0.356} & \textcolor{black}{0.370}\tabularnewline
100 & N/A & 0.233 & 0.188 & 0.208 & 0.282 & \textcolor{black}{0.354} & \textcolor{black}{0.374}\tabularnewline
\hline
\end{tabular}
\par\end{centering}

\end{table*}

\begin{table*}[!t]
\caption{Category-level MRR1 with respect to different numbers of CCA/MVE components
(Using lyrics as query)}

\label{tab:mrr1tclass_dim} 

\begin{centering}
\begin{tabular}{c|c|c|c|c|c|c|c}
\hline
\#CCA/MVE & Spotify-CCA & PretrainCNN-CCA & Spotify-MVE & PretrainCNN-MVE & Spotify-DCCA & Pretrain-DCCA & JointTrain-DCCA\tabularnewline
\hline
10 & 0.178 & 0.170 & 0.246 & 0.277 & 0.256 & 0.314 & 0.366\tabularnewline
20 & 0.176 & 0.188 & 0.249 & 0.304 & 0.294 & 0.344 & 0.368\tabularnewline
30 & 0.179 & 0.198 & 0.222 & 0.273 & 0.305 & 0.351 & 0.372\tabularnewline
40 & 0.185 & 0.208 & 0.204 & 0.235 & 0.307 & 0.358 & 0.365\tabularnewline
50 & 0.191 & 0.220 & 0.199 & 0.228 & 0.306 & 0.355 & 0.373\tabularnewline
60 & 0.190 & 0.223 & 0.195 & 0.221 & 0.302 & 0.356 & 0.374\tabularnewline
70 & N/A & 0.231 & 0.190 & 0.208 & 0.298 & \textcolor{black}{0.360} & \textcolor{black}{0.371}\tabularnewline
80 & N/A & 0.236 & 0.191 & 0.205 & 0.290 & \textcolor{black}{0.354} & \textcolor{black}{0.370}\tabularnewline
90 & N/A & 0.237 & 0.186 & 0.194 & 0.288 & \textcolor{black}{0.356} & \textcolor{black}{0.369}\tabularnewline
100 & N/A & 0.238 & 0.180 & 0.203 & 0.280 & \textcolor{black}{0.355} & \textcolor{black}{0.375}\tabularnewline
\hline
\end{tabular}
\par\end{centering}

\end{table*}

\subsection{Impact of the number of training samples}

Here we investigate the impact of the number of training samples,
by adjusting the percentage of samples for training from 20\% to 80\%.
The percentage of samples for the retrieval test remains 20\%, and
the number of training samples is chosen in such a way that there
are the same number of songs per mood category.

\begin{figure}[tb]
\centering

\includegraphics[width=8cm]{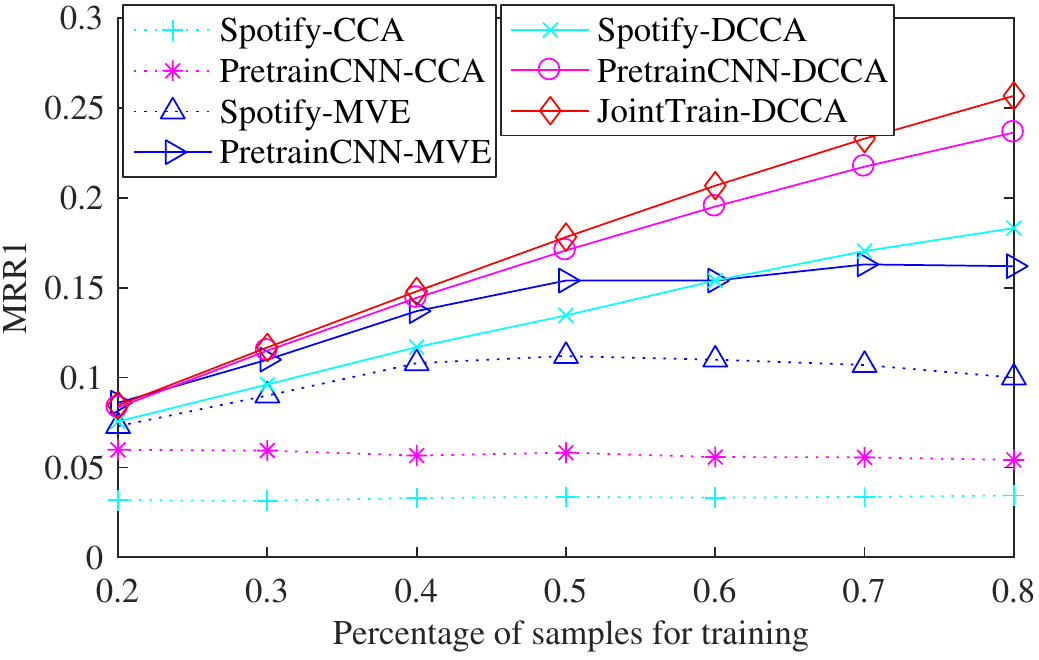}\caption{\label{fig:mrr1a_ratio} Instance-level MRR1 under different percentages
of training samples (Using audio as query to search text lyrics, \#CCA-component=30,
20\% for testing)}
\end{figure}

\begin{figure}
\centering

\includegraphics[width=8cm]{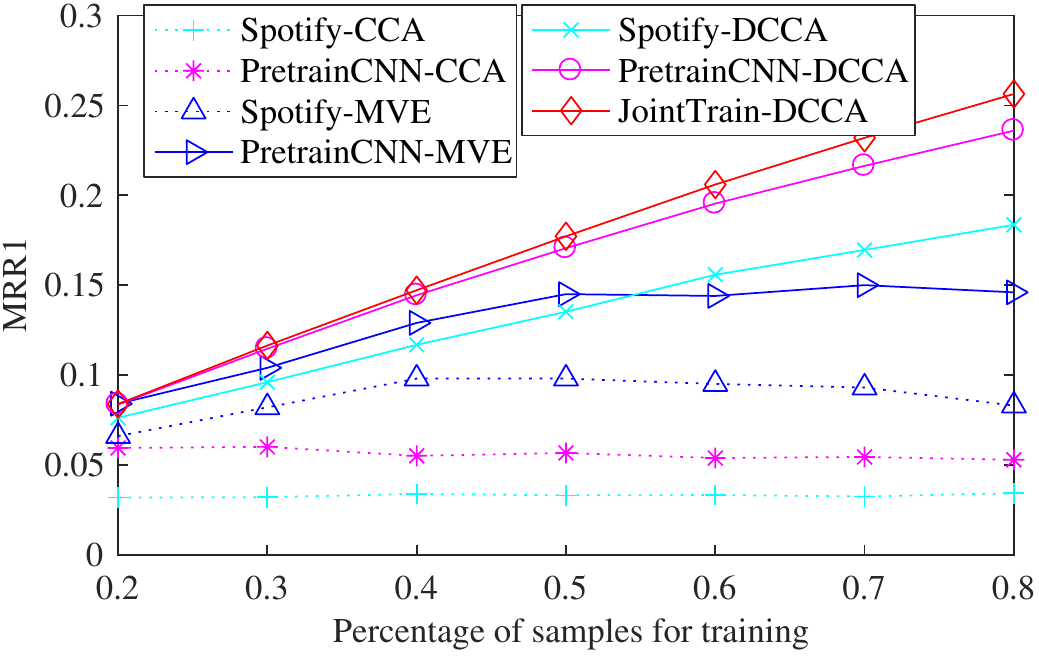}\caption{\label{fig:mrr1t_ratio} Instance-level MRR1 under different percentages
of training samples (Using text lyrics as query to search audio signal,
\#CCA-component=30, 20\% for testing)}
\end{figure}

Fig.~\ref{fig:mrr1a_ratio} and Fig.~\ref{fig:mrr1t_ratio} show
the MRR1 results in the instance-level retrieval. Spotify-CCA and
PretrainCNN-CCA do not benefit from the increase of the training samples.
Spotify-MVE and PretrainCNN-MVE benefits a little. In comparison,
when DCCA is used, the increase of training samples enables the system
to learn more diverse aspect of audio/lyric features, and the MRR1
performance almost linearly increases. In the future, we will try
to crawl more data for training a better model to improve the retrieval
performance.

The MRR1 result, with lyrics as query to search audio, as shown in
Fig.~\ref{fig:mrr1t_ratio}, has a similar trend as that in Fig.
\ref{fig:mrr1a_ratio}.

\begin{figure}[tb]
\centering

\includegraphics[width=8cm]{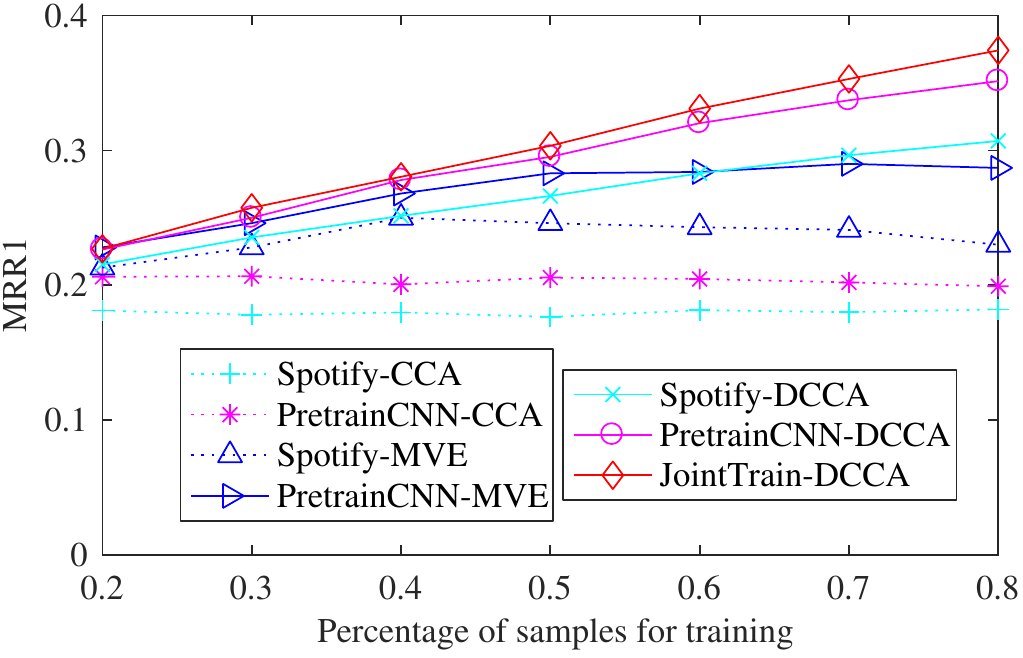}\caption{\label{fig:mrr1aclass_ratio} Category-level MRR1 under different
percentages of training samples (Using audio signal as query to search
text lyrics, \#CCA-component=30, 20\% for testing) }
\end{figure}

\begin{figure}[tb]
\centering

\includegraphics[width=8cm]{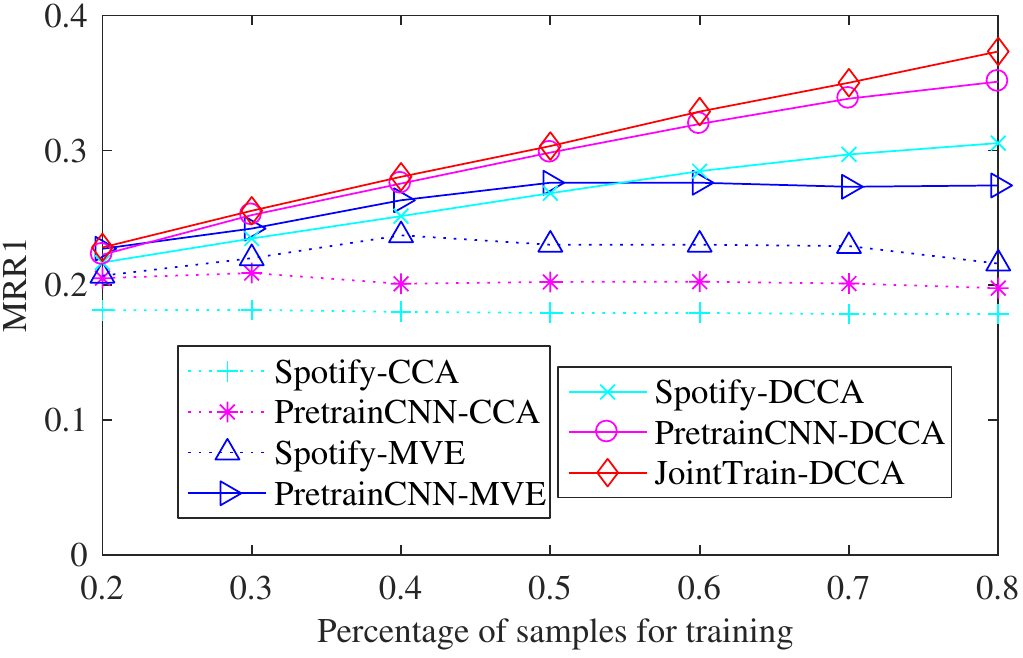}\caption{\label{fig:mrr1tclass_ratio} Category-level MRR1 under different
percentages of training samples (Using text lyrics as query to search
audio signal, \#CCA-component=30, 20\% for testing)}
\end{figure}

Fig.~\ref{fig:mrr1aclass_ratio} and Fig.~\ref{fig:mrr1tclass_ratio}
show the MRR1 results when the retrieval is performed in the category
level. This has a similar trend as the result of instance-level retrieval.

\section{Conclusion}

\label{sec:Conclusion}

Understanding the correlation between different music modalities is
very useful for content-based cross-modal music retrieval and recommendation.
Audio and lyrics are most interesting aspects for storytelling music
theme and events. In this paper, a deep correlation learning between
audio and lyrics is proposed to understand music audio and lyrics.
This is the first research for deep cross-modal correlation learning
between audio and lyrics. Some efforts are made to give a deep study
on i) deep models for processing audio branch are investigated such
as pre-trained CNN with or without being followed by fully-connected
layers. ii) An end-to-end convolutional DCCA is further proposed to
learn correlation between audio and lyrics where feature extraction
and correlation learning are simultaneously performed and joint representation
is learned by considering temporal structures. iii) Extensive evaluations
show the effectiveness of the proposed deep correlation learning architecture
where convolutional DCCA performs best when considering retrieval
accuracy and converging time. More importantly, we apply our architecture
to the bidirectional retrieval between audio and lyrics, e.g., searching
lyrics with audio and vice versa. Cross-modal retrieval performance
is reported at instance level and mood category level.

This work mainly pays attention to studying deep models for processing
music audio while keeping pre-trained Doc2vec for processing lyrics
in correlation learning. We are collecting more audio-lyrics pairs
to further improve the retrieval performance, and will integrate different
music modality data to implement personalized music recommendation.
In the future work, we will investigate some deep models for processing
lyrics branch. Lyrics contain a hierarchical composition such as verse,
chorus, bridge. We will extend our deep architecture to complement
musical composition (given music audio) where Long Short Term Memory
(LSTM) will be applied for learning lyrics dependencies.

\bibliographystyle{IEEEtran}
\bibliography{IEEEabrv,mybibfile}

\end{document}